\documentstyle [preprint,aps,epic,eepic,axodraw]{revtex}
\begin{document}
\def\fnote#1#2{
\begingroup\def\thefootnote{#1}\footnote{#2}\addtocounter{footnote}{-1}
\endgroup}
\def\dslash{\not{\hbox{\kern-2pt $\partial$}}}
\def\eslash{\not{\hbox{\kern-2pt $\epsilon$}}}
\def\Dslash{\not{\hbox{\kern-4pt $D$}}}
\def\Aslash{\not{\hbox{\kern-4pt $A$}}}
\def\Qslash{\not{\hbox{\kern-4pt $Q$}}}
\def\pslash{\not{\hbox{\kern-2.3pt $p$}}}
\def\kslash{\not{\hbox{\kern-2.3pt $k$}}}
\def\qslash{\not{\hbox{\kern-2.3pt $q$}}}
\def\np#1{{\sl Nucl.~Phys.~\bf B#1}}
\def\pl#1{{\sl Phys.~Lett.~\bf B#1}}
\def\pr#1{{\sl Phys.~Rev.~\bf D#1}}
\def\prl#1{{\sl Phys.~Rev.~Lett.~\bf #1}}
\def\cpc#1{{\sl Comp.~Phys.~Comm.~\bf #1}}
\def\anp#1{{\sl Ann.~Phys.~(NY) \bf #1}}
\def\etal{{\em et al.}}
\def\half{{\textstyle{1\over2}}}
\def\be{\begin{equation}}
\def\ee{\end{equation}}
\def\ba{\begin{array}}
\def\ea{\end{array}}
\def\tr{{\rm tr}}
\title{On final-state effects in t\=t production at threshold\thanks{Research
supported in part by the DoE under grant DE--FG05--91ER40627.}}
\author{George Siopsis}
\address{Department of Physics and Astronomy, \\
The University of Tennessee, Knoxville, TN 37996--1200.\\
{\tt siopsis@panacea.phys.utk.edu}}
\date{April 1995}
\preprint{UTHEP--95--0103}
\maketitle
\begin{abstract}
We apply the relativistic Bethe-Salpeter formalism to the calculation of
final-state effects in the production of a t\=t pair at threshold. We find
that final-state rescattering does not affect the momentum distribution of
the t\=t pair to lowest order in the strong coupling constant. This result
correctly extends earlier results based on the non-relativistic
Lippmann-Schwinger equation.
\end{abstract}
\renewcommand\thepage{}\newpage\pagenumbering{arabic}
As was first pointed out by Fadin and Khoze~\cite{rf1a}, the structure of the
t\=t threshold region is very interesting. The cross-section is enhanced by
t\=t resonances and depends strongly on the mass of the top quark $m_t$, its
decay width $\Gamma_t$, and the strong coupling constant $\alpha_s(m_t)$.
Unlike the other heavy quarks (b and c), t\=t pairs cannot form narrow
resonances. This is due to the large value of $m_t$,
which implies that the dominant decay mode is the
weak decay $t\to W^+ b$ with a width $\Gamma_t \sim 1$ GeV. Thus $\Gamma_t$
exceeds $\Lambda_{QCD}$ and as a result the t\=t pair decays before it has time
to hadronize, which makes perturbation theory applicable. Since theoretical
predictions are possible, it would be extremely interesting to measure the
cross-section for threshold t\=t production at the next linear $e^+e^-$
collider (NLC).

In the threshold region, the t\=t pair is produced with little kinetic energy.
Therefore, one can employ non-relativistic techniques to calculate the
cross-section. In analogy with the Hydrogen atom, the t\=t pair is (weakly)
bound together by a QCD Coulomb-like effective potential. The cross-section
for t\=t production is related to the Green function, which obeys an
appropriate Lippmann-Schwinger equation, by the optical theorem \cite{rf2}.
Although it is possible to use this method to calculate higher-order
corrections to the cross-section \cite{rf2}, there are subtleties which ought
to be taken into account. As was shown by Kummer and M\"odritsch \cite{rf4},
these problems can be avoided if one employs the relativistic Bethe-Salpeter
formalism, in analogy with the abelian positronium case \cite{rf6}.

In ref.~\cite{ego}, we extended the results of ref.~\cite{rf4} by showing that
electroweak corrections to the decay width of toponium are suppressed by at
least four powers of the strong coupling constant, $\alpha_s$. To this end, we
employed the covariant Lorentz gauge and perturbed around the
solution to the Bethe-Salpeter equation in the instantaneous approximation.
The calculation was manifestly gauge-invariant and correctly took into account
all contributions to the decay rate (Coulomb enhancement and phase-space
reduction effects).

Here, we wish to apply the argument of ref.~\cite{ego} to the calculation of
radiative corrections to the momentum distribution of the t\=t pair produced
at threshold at a future $e^+e^-$ collider (e.g., the NLC). One-loop
corrections to the
{\em total} cross-section of t\=t threshold production have already been shown
to vanish to first-order in $\alpha_s$ by Melnikov and Yakovlev \cite{rf3}, who
employed the non-relativistic potential formalism. By using the relativistic
Bethe-Salpeter formalism, we shall show that there are no $o(\alpha_s)$
corrections to the {\em differential} cross-section either.

\begin{figure}
 \begin{picture}(300,150)(-50,-20)
\Photon(100,50)(180,50) 5 2 \put(120,30){\large $\gamma \;,\; Z^0$}
\Vertex(100,50) 2\Vertex(180,50) 2
\ArrowLine(180,50)(220,65)\put(200,70){\large $t$}
\ArrowLine(180,50)(220,35)\put(200,20){\large $\bar t$}
\Line(260,65)(310,65)\Line(260,35)(310,35)\Vertex(310,65) 2\Vertex(310,35) 2
\Line(310,65)(340,95)\Line(310,35)(340,5)
\Photon(310,65)(350,65) {-3} 2 \Photon(310,35)(350,35) 3 2
\put(360,65){\large $W^+$}\put(360,25){\large $W^-$}
\put(350,90){\large $b$}\put(350,0){\large $\bar b$}
\filltype{shade}\put(240,50){\circle*{50}}
\ArrowLine(0,0)(100,50)\put(0,10){\large $e^-$}
\ArrowLine(0,100)(100,50)\put(0,80){\large $e^+$}\end{picture}
\caption{\em Threshold t\=t production at an $e^+e^-$ collider (NLC).}
\label{fig2}
\end{figure}
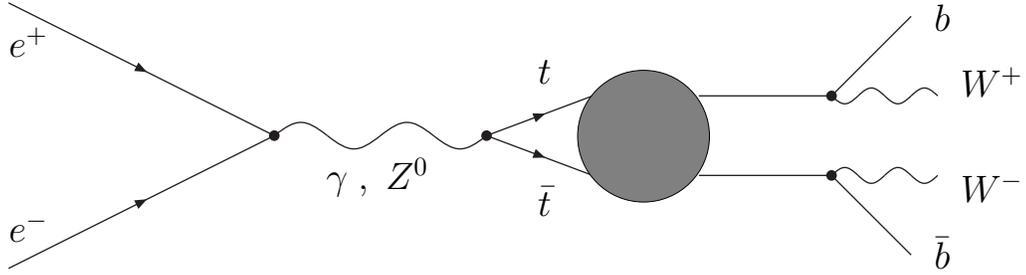

We are interested in one-loop corrections to the process (Fig.~\ref{fig2})
\begin{equation}
  \label{ena}
  e^+ e^- \longrightarrow \gamma\,,\, Z^0 \longrightarrow t\bar t
\longrightarrow W^+ b \; W^- \bar b \;,
\end{equation}
where at threshold the dominant contribution is due to the t\=t bound states
(toponium). To simplify the discussion, we shall ignore t\=t production via
the $Z^0$ boson. We shall also neglect the Higgs boson exchange between t and
\=t, which contributes to the formation of bound states. Our calculation can
easily be extended to accommodate these additional effects.

Let us first review the non-relativistic potential formalism in order to
demonstrate its power as well as its limitations. The t\=t
system at threshold forms short-lived bound states which are eigenstates of the
Hamiltonian
\begin{equation}
  \label{duo}
  H = {{\bf p^2} \over M_t} - {C_F \alpha_s \over r} + 2M_t \;,
\end{equation}
where we have included the effects of the decay in the mass
parameter $M_t =m_t + i\Gamma_t$. The Green function for this Hamiltonian,
\begin{equation}
  \label{tria}
  G({\bf x, x'}; E) = -\sum_n {\psi_n ({\bf x}) \psi_n^\star ({\bf x'}) \over
E-E_n-i\Gamma_n} \;,
\end{equation}
is related to the cross-section of t\=t production by the optical theorem,
\begin{equation}
  \label{tess}
  \sigma (\gamma \rightarrow t\bar t) = {6\pi^2 Q_t^2 \alpha^2 \over m_t^4}
Im \; G({\bf 0,0} ;E)\;,
\end{equation}
where $E=\sqrt s -2m_t$.
This expression for the cross-section includes the contributions of all
ladder diagrams containing an arbitrary number of Coulomb-like gluon exchanges
between the two quarks (t and \=t). This class of diagrams dominates in the
threshold region. By splitting the Hamiltonian (\ref{duo}) in its real and
imaginary parts, $H=H_0+ i\Gamma$, and expanding,
\begin{equation}
  \label{pente}
  {1\over H-E} = {1\over H_0-E}- {1\over H_0-E}i\Gamma {1\over H_0-E} +\dots\;,
\end{equation}
we may write to first-order in the Fermi constant\fnote{\dagger}{The
absorptive part of the Hamiltonian, $\Gamma$, is proportional to $G_F$.}
\begin{equation}
  \label{exi}
  \sigma (\gamma \rightarrow t\bar t) = {3\pi^2 Q_t^2 \alpha^2 \over m_t^4}
\int {d^3 p\over (2\pi)^3} \;|\widetilde G ({\bf p} ;E)|^2\;
\Gamma ({\bf p} ;E)\;.
\end{equation}
The function $\Gamma ({\bf p} ;E)$ represents the decay rate of toponium. To
lowest order, $\Gamma = 2\Gamma_t$. Higher-order corrections are due to
time dilatation and are negligible \cite{rf4,ego}. Thus, the momentum
distribution of the top quark to lowest order is given by
\begin{equation}
  \label{epta}
  {d\sigma \over d|{\bf p}|} = {3Q_t^2 \alpha^2 \over 2m_t^4}{\bf p}^2 \;|
\widetilde G ({\bf p} ;E)|^2 \;.
\end{equation}
Higher-order corrections to the cross-section (\ref{tess}) can be calculated in
this formalism using perturbation theory. However, one needs to be careful in
applying ordinary perturbation theory, because the decay products (e.g., b and
\=b) feel the binding QCD potential and therefore, they cannot be described by
plane waves. To avoid such subtleties, one may use the relativistic
Bethe-Salpeter formalism which is manifestly gauge-invariant.

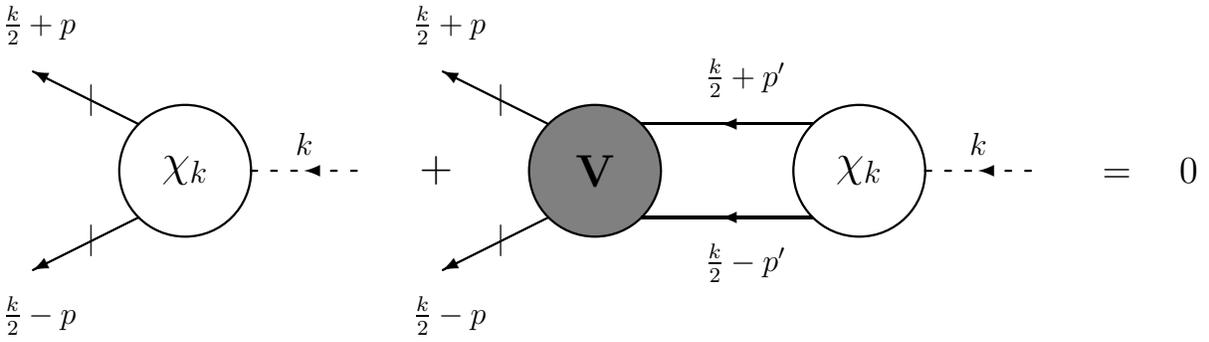
\begin{figure}
\unitlength=1.0pt
\begin{picture}(565,130)(90,680)
\thicklines\filltype{shade}
\put(320.00,765.00){\circle*{50}}
\put(420.00,765.00){\circle{50}}
\put(165.00,765.00){\circle{50}}
\put(165.00,765.00){\makebox(0,0)[cc]{\Large $\chi_k$}}
\put(110.00,820.00){\makebox(0,0)[cc]{${k\over 2} +p$}}
\put(110.00,710.00){\makebox(0,0)[cc]{${k\over 2} -p$}}
\put(260,765){\makebox(0,0)[cc]{\Large $+$}}
\put(265.00,820.00){\makebox(0,0)[cc]{${k\over 2} +p$}}
\put(265.00,710.00){\makebox(0,0)[cc]{${k\over 2} -p$}}
\put(320.00,765.00){\makebox(0,0)[cc]{\Large \bf V}}
\put(377.00,800.00){\makebox(0,0)[cc]{${k\over 2} +p'$}}
\put(377.00,730.00){\makebox(0,0)[cc]{${k\over 2} -p'$}}
\put(420.00,765.00){\makebox(0,0)[cc]{\Large $\chi_k$}}
\put(530,765){\makebox(0,0)[cc]{\large $=\quad 0$}}
\put(337.7,747.3){\line( 1, 0){ 30.5}}
\put(402.3,747.3){\vector( -1, 0){ 34.64}}
\put(337.7,782.7){\line( 1, 0){ 30.5}}
\put(402.3,782.7){\vector( -1, 0){ 34.64}}
\put(147.3,747.3){\vector(-2,-1){ 40}}
\put(129.4,738.4){\makebox(0,0)[cc]{$|$}}
\put(147.3,782.7){\vector(-2, 1){ 40}}
\put(129.4,791.6){\makebox(0,0)[cc]{$|$}}
\put(302.3,747.3){\vector(-2,-1){ 40}}
\put(284.4,738.4){\makebox(0,0)[cc]{$|$}}
\put(302.3,782.7){\vector(-2,1){ 40}}
\put(284.4,791.6){\makebox(0,0)[cc]{$|$}}
\dashline[+30]{3}(190,765)(230,765)
\put(210,765){\vector( -1, 0){ 0}}
\put(210,775){\makebox(0,0)[cc]{$k$}}
\dashline[+30]{3}(445,765)(485,765)
\put(465,765){\vector( -1, 0){ 0}}
\put(465,775){\makebox(0,0)[cc]{$k$}}
\end{picture}
\caption{\em The homogeneous Bethe-Salpeter equation.}
\label{fig1}
\end{figure}

To calculate the differential cross-section, we note that it factorizes into
two parts, describing t\=t production ($e^+e^- \to \gamma \;,\; Z^0 \to t\bar
t$), and scattering and decay of the t\=t pair ($t\bar t \to bW^+\,
\bar b W^-$), respectively. We shall concentrate on the latter part. Let
$k^\mu$ be the total incoming momentum, and $p^\mu$ be the relative momentum
of the t\=t pair. After we integrate over the final-state phase space, we can
use the optical theorem to express the cross-section as
\begin{equation}
  \label{octo}
  {d\sigma \over d|{\bf p}|} \sim Im \, {\cal A}_{t\bar t} (k,p,p') \;,
\end{equation}
where ${\cal A}_{t\bar t}$ is the amplitude for t\=t scattering, and $p^{'\mu}$
is the relative momentum of the final t\=t pair. This four-point amplitude
satisfies the inhomogeneous Bethe-Salpeter equation,
\begin{equation}
  \label{ennia}
\Pi^{(1)} (p_+) \Pi^{(2)} (p_-) {\cal A}_{t\bar t} (k,p,p') =  1+\int {d^4 p''
\over (2\pi )^4} V(p,p'';k) {\cal A}_{t\bar t} (k,p'',p')\;,
\end{equation}
where $\Pi(p)$ is the complete inverse fermion propagator,
\begin{equation}
\Pi (p) = \pslash -M_t -\Sigma (p) \;,
\label{eq4}\end{equation}
and we have defined momenta
\begin{equation} p_\pm = {k\over 2} \pm p \;.\label{eq4a}\end{equation}
$V(p,p';k)$ is a potential function which consists of the two-fermion
irreducible graphs. For our purposes, the mass is complex,
\begin{equation} M_t= m_t + i\Gamma_t \;. \label{eq5} \end{equation}
The solution to Eq.~(\ref{ennia}) has poles which are due to bound states.
Near a pole,
\begin{equation}
  \label{deka}
{\cal A}_{t\bar t} (k,p,p') \sim {i\overline\chi_k(p) \chi_k(p')\over
k^2 -M^2} \;.
\end{equation}
The bound-state wavefunctions $\chi_k(p)$ satisfy the homogeneous
Bethe-Salpeter equation,
\begin{equation}
\Pi^{(1)} (p_+) \Pi^{(2)} (p_-)\chi_k(p) + \int {d^4 p' \over (2\pi )^4}
V(p,p';k) \chi_k (p') =0 \;.
\label{eq3}\end{equation}
which is represented graphically in Fig.~\ref{fig1}.

To lowest order in $\alpha_s$ and neglecting electroweak interactions,
the potential is
\be V_0 (p,p';k) = C_F \, 4\pi\alpha_s \gamma_\mu^{(1)} G^{\mu\nu}
(p-p') \gamma_\nu^{(2)} \;, \label{eq6} \ee
where $C_F=4/3$ is the Casimir operator, and
$G^{\mu\nu} (q)$ is the lowest-order gluon propagator. In the Feynman
gauge (omitting group theory factors),
\be G^{\mu\nu} (q) = {\eta^{\mu\nu} \over q^2 +i\epsilon } \;, \label{eq7} \ee
and the potential $V_0 (p,p';k)$ is independent of $k^\mu$.
At threshold, the quarks move with non-relativistic velocities and the
Bethe-Salpeter equation can be approximated by the non-relativistic
Schr\"odinger equation in momentum space, and then solved. To this end,
we shall work in the total rest frame in which the overall momentum is
$k^\mu = (E, \vec 0)$. In the instantaneous approximation, the potential
becomes
\be V_0^{inst}(p,p';k) = C_F \, 4\pi\alpha_s \gamma_0^{(1)} {1\over
(\vec p-{\vec p\, }')^2} \gamma_0^{(2)} \;. \label{eq6a} \ee
If we integrate over $p^0$, we can write the Bethe-Salpeter equation
(\ref{eq3}) in terms of the wavefunction $\Phi (\vec p) = \int {dp^0 \over
2\pi} \chi (p)$ as
\be (H^{(1)} + H^{(2)} - E) \Phi (\vec p) = \left( \Lambda_+^{(1)}
\Lambda_+^{(2)} - \Lambda_-^{(1)} \Lambda_-^{(2)} \right) C_F \,
4\pi\alpha_s \int {d^3 p\over (2\pi)^3} {1\over (\vec p-{\vec p\, }')^2}
\Phi ({\vec p\, }') \;, \label{eq6b}\ee
where $H$ is the Dirac Hamiltonian and $\Lambda_+$ ($\Lambda_-$) is the
projection operator onto positive (negative) energy states.
In the non-relativistic limit, this reduces to the Schr\"odinger equation
in momentum space
\be \left( {{\vec p\, }^2 \over M_t} +2M_t- E \right) \Phi (\vec p) = C_F \,
4\pi\alpha_s \int {d^3 p\over (2\pi)^3} {1\over (\vec p-{\vec p\, }')^2}
\Phi ({\vec p\, }') \;. \label{eq6c}\ee
Thus, we obtain the energy levels
\be E_n = 2M_t - {M_t C_F^2\alpha_s^2 \over 4n^2} + o(\alpha_s^4 ) \;,
\label{eq8} \ee
which are the Bohr levels of the Coulomb-like QCD potential (\ref{eq6a}).
Therefore, the first-order QCD correction to the decay rate of toponium is
\be \Gamma_{t\bar t} = 2\Gamma_t \left( 1- {C_F^2\alpha_s^2 \over 8n^2}\right)
\;, \label{eq9} \ee
which may be attributed to time dilatation~\cite{rf4}.
The spherically symmetric $S=0$ states are given by
\be \Phi_n (\vec p) = (M_tC_F \alpha_s)^{-3/2} {{\cal L}_n (n^2y) \over
(1+n^2y)^{n+1}}\;\;, \;\;
y={4{\vec p \, }^2 \over M_t^2 C_F^2 \alpha_s^2} \,, \label{eq9a}\ee
where ${\cal L}_n$ is a polynomial of order $n-1$ related to the Laguerre
polynomials. For $n=1$, we have ${\cal L}_1= 16\sqrt{2\pi}$.

\begin{figure}
\begin{center}
\begin{picture}(300,120)(0,-20)
\PhotonArc(-10,50)(30,0,180) 4 4
\Line(60,50)(-80,50)
\Line(60,0)(-80,0)
\Vertex(-40,50) 2
\Vertex(-10,50) 2
\Vertex(-10,0) 2
\Gluon(-10,50)(-10,0) {-4} 3
\Vertex(20,50) 2
\put(-70,60){\makebox(0,0)[cc]{$t$}}
\put(-70,10){\makebox(0,0)[cc]{$\bar t$}}
\put(-10,95){\makebox(0,0)[cc]{$W\, ,\, \Phi$}}
\Line(80,50)(220,50)
\Line(80,0)(220,0)
\Gluon(110,50)(110,0) {-4} 3
\PhotonArc(165,50)(30,0,180) 4 4
\Vertex(110,50) 2
\Vertex(135,50) 2
\Vertex(195,50) 2
\Vertex(110,0) 2
\put(90,60){\makebox(0,0)[cc]{$t$}}
\put(90,10){\makebox(0,0)[cc]{$\bar t$}}
\put(165,95){\makebox(0,0)[cc]{$W\, ,\, \Phi$}}
\put(165,40){\makebox(0,0)[cc]{$b$}}
\Line(240,50)(380,50)
\Line(240,0)(380,0)
\Gluon(350,50)(350,0) {-4} 3
\Vertex(350,50) 2
\Vertex(350,0) 2
\Vertex(265,50) 2
\Vertex(325,50) 2
\PhotonArc(295,50)(30,0,180) 4 4
\put(295,40){\makebox(0,0)[cc]{$b$}}
\put(250,60){\makebox(0,0)[cc]{$t$}}
\put(250,10){\makebox(0,0)[cc]{$\bar t$}}
\put(295,95){\makebox(0,0)[cc]{$W\, ,\, \Phi$}}
\end{picture}
\end{center}
\caption{\em First-order electroweak corrections to t\=t production.}
\label{fig3}
\end{figure}
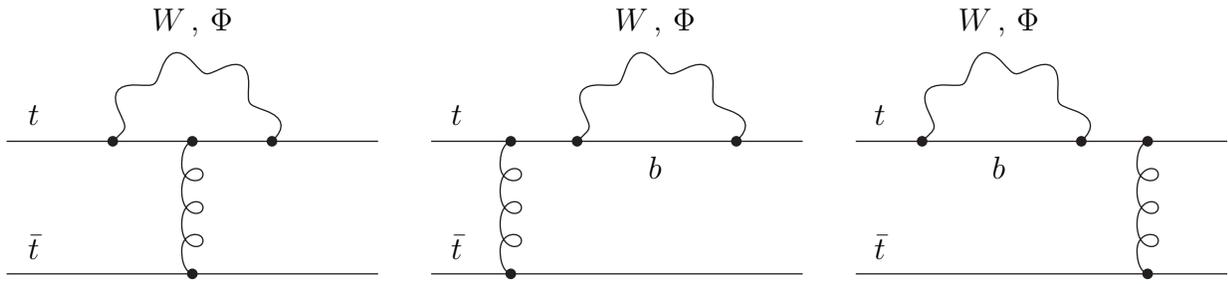

\begin{figure}
\begin{center}
\begin{picture}(300,100)(0,-20)
\Line(60,50)(-80,50)
\Line(60,0)(-80,0)
\Vertex(-10,50) 2
\Gluon(-10,50)(40,0) {-4} 4
\Vertex(40,0) 2\Vertex(10,50) 2\Vertex(10,0) 2
\Photon(10,50)(60,80) 3 2
\Photon(10,0)(60,-30) {-3} 2
\put(-70,60){\makebox(0,0)[cc]{$t$}}
\put(-70,10){\makebox(0,0)[cc]{$\bar t$}}
\Line(80,50)(220,50)
\Line(80,0)(220,0)
\Gluon(200,50)(150,0) {-4} 4
\Vertex(200,50) 2
\Vertex(150,0) 2
\Vertex(170,50) 2
\Vertex(170,0) 2
\put(90,60){\makebox(0,0)[cc]{$t$}}
\put(90,10){\makebox(0,0)[cc]{$\bar t$}}
\Photon(170,50)(220,80) 3 2
\Photon(170,0)(220,-30) {-3} 2
\Line(240,50)(380,50)
\Line(240,0)(380,0)
\Gluon(350,50)(350,0) {-4} 3
\Vertex(350,50) 2
\Vertex(350,0) 2
\Vertex(330,50) 2
\Vertex(330,0) 2
\Photon(330,50)(380,80) 3 2
\Photon(330,0)(380,-30) {-3} 2
\put(250,60){\makebox(0,0)[cc]{$t$}}
\put(250,10){\makebox(0,0)[cc]{$\bar t$}}
\end{picture}
\end{center}
\caption{\em Final-state t\=t rescattering.}
\label{fig4}
\end{figure}
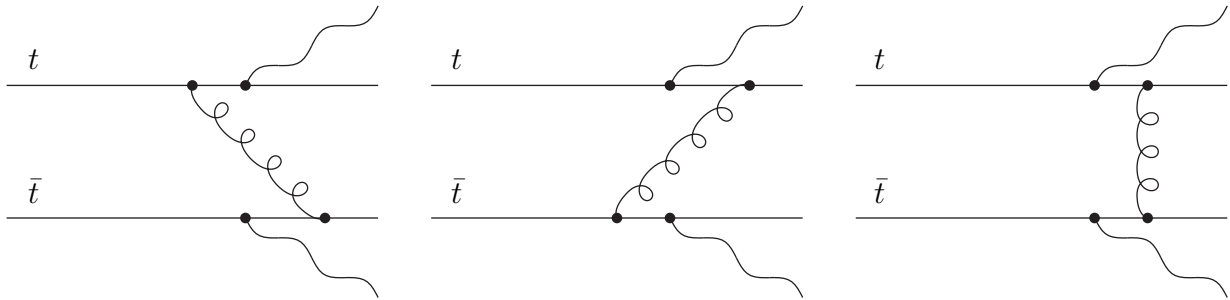

To discuss final-state corrections, we need to consider the diagrams in
Fig.~\ref{fig3}.
They give first-order corrections to the differential cross-section for t\=t
decay, according to the unitarity theorem. This may be seen by cutting these
diagrams to produce the graphs of Fig.~\ref{fig4}. In the positronium case,
these loop
graphs represent magnetic moment effects and contribute to a $o(\alpha^5)$
shift in the energy levels (poles). We have shown that there is no $W$ boson
contribution to the color magnetic moment of the top quark \cite{ego}. This is
due to the
fact that $W$ only couples to a left-handed current. We shall sketch the proof
of this statement for completeness.

The energy level shifts can be calculated by perturbing around the
solution to the Schr\"odinger equation (\ref{eq6c}). The potential to be
treated perturbatively is $V-V_0^{inst}$. There is also a contribution from
the disconnected diagrams which are due to the self-energy terms in the
fermion propagators (Eq.~(\ref{eq4})), but they can be absorbed in the
potential if we make use of the Schr\"odinger equation. Thus, according to
the Bethe-Salpeter formalism \cite{rf6}, the first-order energy level
shift is
\be \Delta E_n = \langle \Phi_n |\, D_k (p)\, (H^{(1)} +H^{(2)}-E_n)\,
(V-V_0^{inst})\, (H^{(1)} +H^{(2)}-E_n)\, D_k (p)\, | \Phi_n \rangle \;,
\label{eq6d}\ee
where the inner product involves an integral over the four-momentum. $H$ is
the Dirac Hamiltonian, and
$D_k$ is the product of two free fermion propagators
({\em cf.}~Eq.~(\ref{eq3})), which can be expressed
in terms of the projection operators $\Lambda_\pm$ as
\be D_E (p) = \sum_{\pm \pm} {\Lambda_\pm^{(1)}\Lambda_\pm^{(2)} \over
[E/2 + p^0 \pm (E_p -i\epsilon )][E/2 - p^0 \pm (E_p -i\epsilon )]} \;,
\label{eq6e}\ee
where $E_p = \sqrt{{\vec p \, }^2 +m_t^2}$ is the energy of the quark on the
mass shell.

To lowest order, the
potential is $V_0-V_0^{inst}$ (Eqs.~(\ref{eq6}) and (\ref{eq6a})). This is
analogous to the positronium case, and produces an $o(\alpha_s^4)$ shift in
the energy levels. The first-order electroweak correction is
\be V_1 (p,p';k) = 4\pi C_F \alpha_s\alpha_W\bigg(
\Lambda_\mu^{(1)} (p_+,p_+') G^{\mu\nu}
(p-p') \gamma_\nu^{(2)} + \gamma_\mu^{(1)}G^{\mu\nu}
(p-p')\Lambda_\nu^{(2)}(p_-,p_-')\bigg) \;, \label{eq12}\ee
where $p_\pm = k/2 \pm p$, $p_\pm' = k/2 \pm p'$, and we have
made explicit the electroweak coupling constant $\alpha_W \sim G_FM_W^2$,
where $G_F$ is the Fermi constant and $M_W$ is the mass of the $W$ boson.
The vertex function $\Lambda_\mu (p,p')$ consists of the diagrams shown in
fig.~\ref{fig3}. It is guaranteed to give a gauge invariant contribution by
the Ward identity satisfied by the one-particle irreducible function,
\be (p-p')^\mu \Gamma_\mu (p,p') = \Pi (p) - \Pi (p') \;. \ee
Since we are only interested in first-order corrections, we may replace $M_t$
by its real part $m_t$.
The contribution of $V_1$ to the energy level shift (Eq.~(\ref{eq6d})) can then
be written as
$$\Delta E_n^W = {C_F^2 \alpha_s^2 \alpha_W \over 16m_t} \int {d^4 p \over
(2\pi )^4} {d^4 p' \over (2\pi )^4} {\eta^{\mu\nu} \over (p-p')^2 + i\epsilon}
\; {{\cal L}_n (n^2y) \over (1+n^2y)^n}\;
{{\cal L}_n (n^2y') \over (1+n^2y')^n} \hskip 1in$$
\be \hskip 1in\times \left\langle D_k (p') \bigg(
\Lambda_\mu^{(1)} (p_+,p_+') \gamma_\nu^{(2)} + \gamma_\mu^{(1)}
\Lambda_\nu^{(2)}(p_-,p_-')\bigg) D_k (p) \right\rangle \;, \label{nn1}
\ee
where $y=4{\vec p \, }^2 / m_t^2 C_F^2 \alpha_s^2$ and
$y'=4{\vec p \, }'^2 / m_t^2 C_F^2 \alpha_s^2$. A simple scaling argument shows
that the lowest-order contribution to the integral comes from the small
three-momentum region. Momentum insertions contribute additional powers of
$\alpha_s$. At low momentum transfer, the three-point vertex $\Lambda_\mu$ may
be written in general as
\be \Lambda_\mu (p,p') = k^2 {\cal F}_1 (k^2) + \sigma_{\mu\nu} k^\nu
{\cal F}_2 (k^2) \;, \ee
where $k=p-p'$, and $\sigma_{\mu\nu}={i\over 2}[\gamma_\mu \,,\,\gamma_\nu ]$.
The form factors ${\cal F}_1$ and ${\cal F}_2$ are regular
as $k^2 \to 0$. In the positronium case, ${\cal F}_2$ gives an $o(\alpha^5)$
contribution to the energy level shift, and is due to the magnetic moment
interaction. In our case, we need to multiply the gamma matrices by the
projection operator
${1\over 2} (1-\gamma_5)$, due to parity violation of weak interactions.
A straightforward explicit calculation shows that the form factor
${\cal F}_2 (k^2)$ vanishes to lowest order in $k^2$.
It follows that the three-point vertex is
proportional to $(p-p')^2$ (recall that $p_+-p_+'=p_--p_-'=p-p'$).

Having established the leading-order behavior of $\Lambda_\mu$, we can
now estimate the integral in Eq.~(\ref{nn1}). As we just showed, $\Lambda_\mu$
contributes a factor $(p-p')^2$. This factor cancels the gluon propagator.
Then the integral over
$p_0$ and $p_0'$ can be easily done, because of the respective poles in
the operators $D (p)$ and $D(p')$. The resulting expression contains six
three-momentum factors implying that the integral is $o(\alpha_s^8)$.
Therefore, the electroweak correction to the decay width is negligible.
Of course, no conclusion can be drawn regarding the exact value of the
electroweak
correction, because such a high order is beyond the scope of first-order
perturbation theory.

Having shown that the poles do not get shifted due to first-order final-state
corrections, we deduce that there are no first-order corrections to the
amplitude either.
Therefore, the total cross-section {\em as well as} the differential
cross-section of
t\=t production and decay do not get corrected to first-order by final-state
rescattering.

In conclusion, we have presented a simple physical argument showing that
final-state threshold effects vanish to first-order in perturbation theory.
This is true for the differential cross-section of t\=t production, which
strengthens a
previous result on the total cross-section by Melnikov and Yakovlev \cite{rf3}.
Our argument was based on the relativistic Bethe-Salpeter formalism. No special
gauge-fixing procedure was required and the calculation was manifestly
gauge-invariant.

\newpage
\vspace{1in}
\newpage
\mbox{}
\vspace{1in}

\end{document}